\documentstyle[aps,prb,twocolumn,psfig]{revtex}

\draft
\begin{document}
\twocolumn[\hsize\textwidth\columnwidth\hsize\csname@twocolumnfalse%
\endcsname

\title{Magnetic properties of (VO)$_2$P$_2$O$_7$: two-plane structure 
and spin-phonon interactions}

%\title{Spin-Phonon Coupling in (VO)$_2$P$_2$O$_7$}

\author{G. S. Uhrig$^1$ and B. Normand$^2$ }

\address{$^1$Institut f\"ur Theoretische Physik,
Universit\"at zu K\"oln, D-50937 K\"oln, Germany.}

\address{$^2$Theoretische Physik III, Universit\"at Augsburg, 
D-86135 Augsburg, Germany.}

\date{\today}

\maketitle

\begin{abstract}

Detailed experiments on single-crystal (VO)$_2$P$_2$O$_7$ continue 
to reveal new and unexpected features. We show that a model composed 
of two, independent planes of spin chains with frustrated magnetic 
coupling is consistent with nuclear magnetic resonance and inelastic 
neutron scattering measurements. The pivotal role of PO$_4$ groups in 
mediating intrachain exchange interactions explains both the presence 
of two chain types and their extreme sensitivity to certain lattice 
vibrations, which results in the strong magnetoelastic coupling 
observed by light scattering. We compute the respective modifications 
of the spin and phonon dynamics due to this coupling, and illustrate 
their observable consequences on the phonon frequencies, magnon 
dispersions, static susceptibility and specific heat.

\end{abstract}

]

\section{Introduction}

Vanadyl pyrophosphate (VOPO) presents a quantum magnetic system 
of S = ${\textstyle \frac{1}{2}}$ V$^{4+}$ ions, whose antiferromagnetic 
(AF) Heisenberg superexchange interactions give a singlet ground state, 
with spin gap $\Delta = 3.1$meV.\cite{rjjgj} While early powder 
susceptibility and inelastic neutron scattering (INS) experiments were 
consistent with both spin-ladder and alternating-chain models,\cite{rbr}
the first INS measurements on aligned single crystals\cite{rgntsb} 
confirmed the magnetic structure to be of alternating-chain conformation. 
In conjunction with observations on the structurally related material 
VODPO$_4 . {\textstyle \frac{1}{2}}$D$_2$O,\cite{rtngbt} the following 
picture was verified: the strongest exchange path ($J_1$) is the double 
V-O-P-O-V link through two phosphate groups along ${\hat b}$; the next 
strongest ($J_2$) is the double V-O-V link between edge-sharing VO$_5$ 
square pyramids, also along $\hat{b}$; the structurally dimerized 
chains have coupling ratio $\lambda = J_2/J_1 \simeq 0.8$.\cite{rbrt}
The V-O-V bond along $\hat{a}$ is very weak. These results are consistent 
with the single electron on V$^{4+}$ occuping the $d_{xy}$ orbital 
($bc$ plane). 

The same experiment\cite{rgntsb,rgnbs} obtained detailed measurements of a 
second, low-lying, triplet excitation with gap 5.7meV and significant 
intensity over at least half of the Brillouin zone, while in addition 
the coupling between dimerized chains was found to be weakly 
ferromagnetic (FM). Interpretation of the latter features was 
offered\cite{run} in terms of a frustrated coupling of the dimerized 
chains, via the long but presumably not insignificant V-O-P-O-P-O-V 
pathway, which would promote a triplet, two-magnon bound state. Although 
VOPO has a complicated structure,\cite{rnhs} which contains 8 V atoms 
per unit cell, the almost identical interatomic distances $d_{VV}$ 
suggested a model\cite{run} treating all dimers as identical magnetic 
unit cells. 

Since the proposal of this picture, three important, additional 
experiments have been conducted on VOPO single crystals, namely 
further inelastic neutron scattering (INS) measurements, Nuclear 
Magnetic Resonance (NMR) studies and Raman light scattering. Each 
has offered new, and sometimes surprising, additional information 
concerning the physics of VOPO. The purpose of this work is to review 
these results and provide a new, minimal model which encapsulates all 
of the observed effects. We proceed in Sec. II with a discussion of 
INS and NMR results which lead to a two-plane description. In Sec. III 
we consider the strong magnetoelastic coupling observed by Raman 
scattering, and present a theoretical treatment of the coupled 
magnon-phonon system by the flow-equation method. The consequences of 
this coupling are illustrated for a variety of experimentally 
measureable quantities in Sec. IV. Sec. V contains a summary and 
conclusions. 

\section{NMR and bilayer structure}

Contrary to the expectation of magnetically identical V$^{4+}$ 
ions, NMR measurements of the Knight Shift and relaxation rate 
$1/T_1$\cite{rkmyu} give clear evidence for two distinct species 
of $^{31}$P and $^{51}$V nuclei in VOPO. While results for both 
atomic species were not fully consistent, the presence of separate 
magnetic environments is quite unambiguous. An interpretation 
requires two distinct planes of coupled, dimerized chains, with 
different exchange constants $(J_1, J_2)_{A,B}$, resulting in 
different spin gaps $\Delta_{0A}$ and $\Delta_{0B}$. 

This deduction is consistent with the structure,\cite{rnhs} where 8 
inequivalent V atoms occupy 2 dimers per chain and 2 chains per cell: 
despite the very small differences in V-V separations, these dimer 
units are clearly not identical. The very weak $c$-axis coupling, 
observed in the INS dispersion, acts to isolate the differing planes 
of coupled chains. The two triplet branches measured by INS have the 
simple interpretation of two one-magnon dispersion curves, one from 
each plane: this explains why their intensities have very similar 
${\bf q}$-dependences.\cite{rnpc} The reported gaps $\Delta_{0B} = 
35$K and $\Delta_{0A} = 68$K\cite{rkmyu} are in excellent agreement 
with the results of Ref.~\onlinecite{rgntsb}. 

To answer the question of why the $J$ values in the two planes are 
so different, we note that the previous treatment of all dimers as 
identical magnetic unit cells was based on the fact that differences 
in $d_{VV}$ are ${\cal O}(10^{-3}$) of the unit cell dimensions. However, 
inspection of the structural parameters\cite{rnhs} reveals that 
differences in the locations of P atoms, and thus in the interatomic 
P-V distances $d_{PV}$, are ${\cal O}(10^{-2}$). The PO$_4$ group is the 
essential element mediating the dominant superexchange path V-O-P-O-V. 
While quantitative accuracy is still lacking in {\it ab initio} 
superexchange calculations, particularly for V systems,\cite{rmsbnpam} 
the interaction magnitude is well known to have a very strong 
dependence on both interatomic spacings and bond angles. Thus one may 
expect a strong difference between values $J_{1A}$ and $J_{1B}$, but 
similar values of $J_{2A}$ and $J_{2B}$. 

\begin{figure}[hp]
\centerline{\psfig{figure=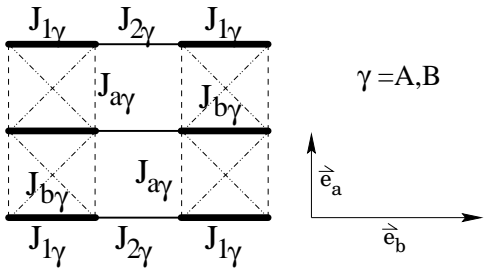,height=3.5cm,angle=0}}
\medskip
\caption{\label{fscheme} 
Schematic representation of the VOPO system; A and B denote 
the two decoupled types of plane.}
\end{figure}

Retaining the basic framework of Ref.~\onlinecite{run}, we consider thus 
a model of two independent types of planes of alternating chains 
(Fig.~\ref{fscheme}).
We begin by fitting the two independent magnon branches\cite{rgntsb} to 
determine the appropriate, effective superexchange constants. This is 
performed within a ``static model'', by which is meant one with 
temperature-independent effective interactions, in contrast with the 
situation in Sec. IV where we will consider a model with phonon dynamics. 
The magnons are described by mobile triplet excitations,\cite{ru} and the 
fitting procedure differs from that presented\cite{run} for the 
single-plane model in two minor respects. First, as mentioned in that 
work, we fit the square of each dispersion curve, 
\begin{equation} 
\omega_{\gamma}^2 ({\bf q}) = J_{1\gamma}^2 \sum_{ij} u_{ij\gamma} 
\cos (i q_y) \cos (j q_x), 
\label{esdr}
\end{equation}
because this smoother quantity gives a superior fit to the same order of 
expansion. The parameters $u_{ij}$, for $\gamma = A,B$, are determined 
to third order from the quantities $t_{ij}$ in Eq. (1) of 
Ref.~\onlinecite{run} as\cite{fn}
\begin{eqnarray}
u_{00} & = & 1 + 7 \lambda^3 / 32 + 2 \mu_{-}^2 + 3 \mu_{-}^2 \mu_{+} 
/ 4 - 2 \lambda \alpha \label{epc} \nonumber \\ & & \;\; + 2 \alpha^2 - 
(15 \lambda^2 \alpha + 6 \lambda \alpha^2 - 12 \alpha^3) / 16 \nonumber 
\\ u_{10} & = & - \lambda - \lambda^2 / 2 + \mu_{-}^2 \alpha / 2 
+ 2 \alpha + 5 \lambda^3 / 32 \nonumber \\ & & \;\; - 7 \lambda^2 
\alpha / 16 - 9 \lambda \alpha^2 / 8 + 3 \alpha^3 / 4 \nonumber \\ 
u_{01} & = & 2 \mu_{-} + 3 \mu_{-}^3 / 4 - \lambda^2 \mu_{-} \\ 
u_{20} & = & (\lambda^3 - 2 \lambda^2 \alpha + 4 \lambda \alpha^2 - 8 
\alpha^3) / 16 \nonumber \\ u_{02} & = & - \mu_{-}^2 \mu_{+} / 2 
\nonumber \\ u_{11} & = & - 3 \lambda^2 \mu_{-} / 4 - \alpha^2 \mu_{-} 
+ \lambda \mu_{-} \mu_{+} / 2 - \alpha \mu_{-} \mu_{+} . 
\nonumber 
\end{eqnarray}
Here $\lambda = J_2/J_1$ (Fig.~\ref{fscheme}), $\mu_+ = (J_a + J_b)/J_1$ and 
$\mu_- = (J_a - J_b)/J_1$. The transverse part of the dispersion 
$\omega_{\gamma} ({\bf q})$ is very sensitive to the difference 
$\mu_{-\gamma}$ of the interchain exchange constants, but rather 
insensitive to their sum $\mu_{+\gamma}$. The parameter $\alpha = 
J_f/J_1$ denotes a frustrating, next-neighbor interaction ($J_f$) along 
the spin chains. It is not present in our minimal model (Fig.~\ref{fscheme}) 
and is not used in the fits presented in this section, but is included here 
for reference from Sec. IV, where such a term is generated within the 
dynamical model. The second difference from Ref.~\onlinecite{run} 
is that we do not determine $\mu_{+\gamma}$ from the condition on the Curie 
temperature deduced from the magnetic susceptibility, as this procedure 
may no longer be applicable in the presence of phonons coupled strongly 
to the spin subsystem (below). Instead we deduce the values 
$\mu_{+\gamma}$ from the maximum $\chi_{\rm max}$ of the measured 
suseptibility curve.\cite{rpbaswll} 

\begin{figure}[hp]
\centerline{\psfig{figure=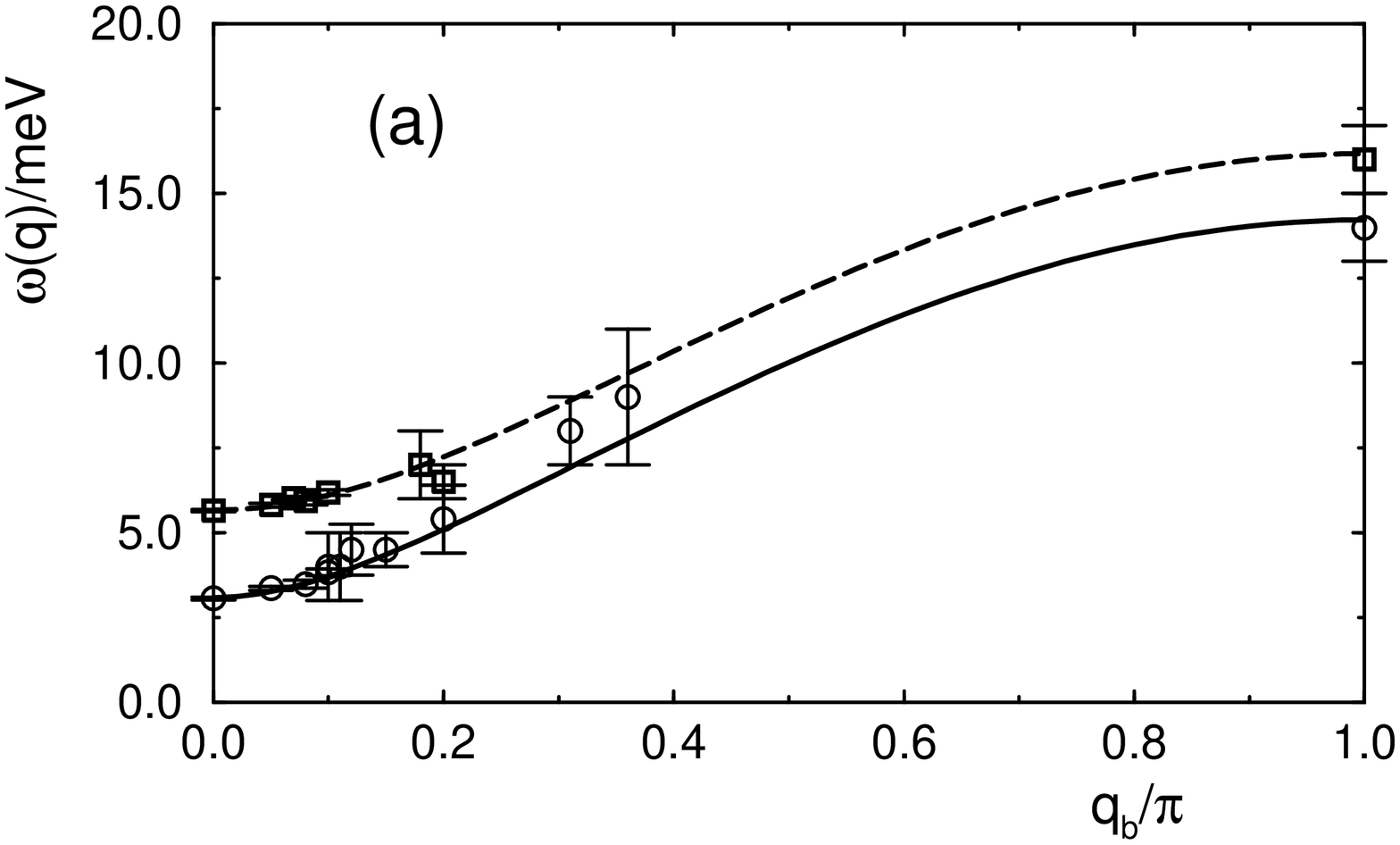,height=5cm,angle=270}}
\centerline{\psfig{figure=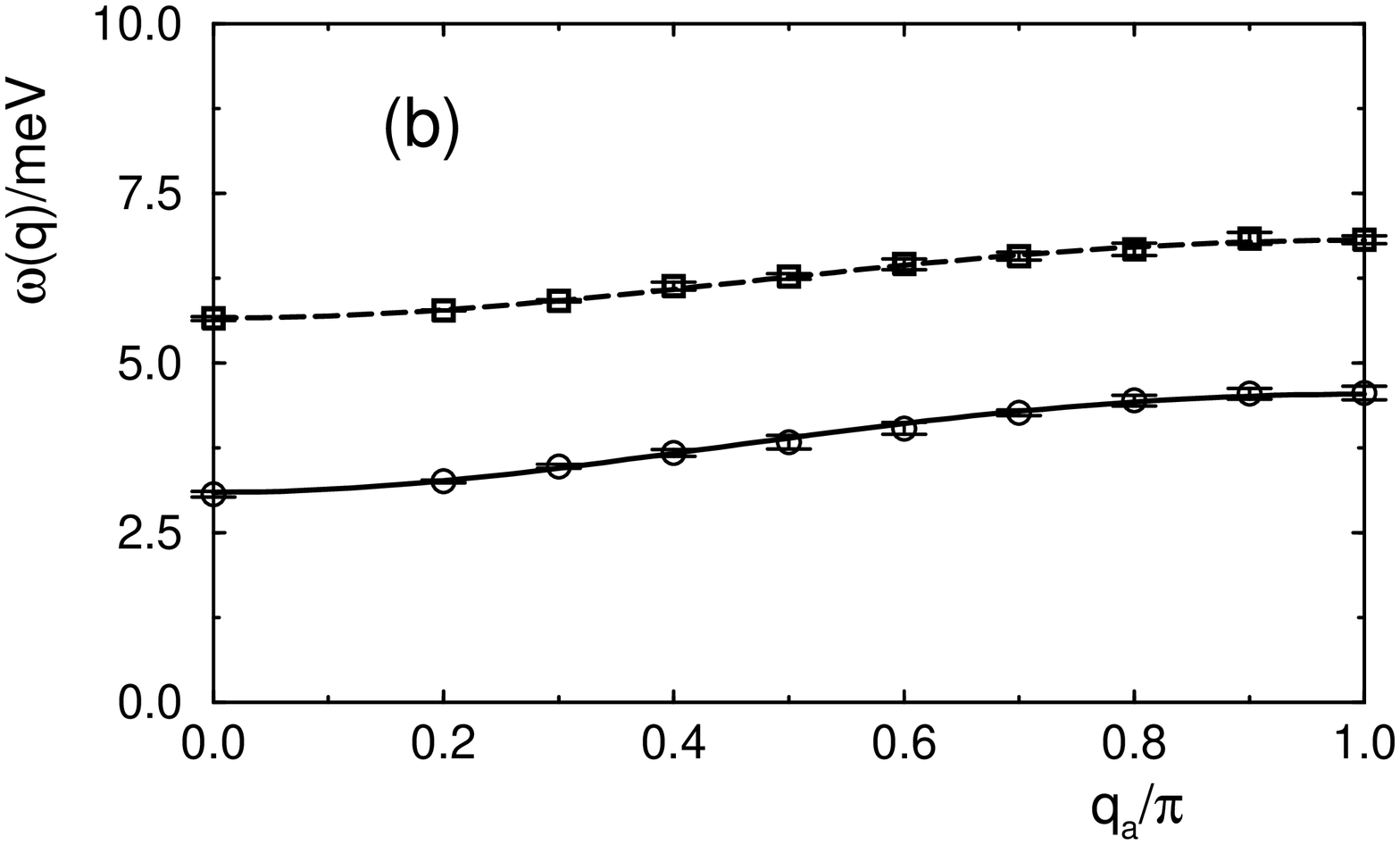,height=5cm,angle=270}}
\medskip
\caption{\label{fdisp}
Fitted dispersion relations for the two elementary magnons, 
(a) in chain direction at $q_a = 0$ and (b) transverse to chains at 
$q_b = 0$. Data from Refs.~\protect\onlinecite{rgntsb} and 
\protect\onlinecite{rspc}. }
\end{figure}

The fit (Fig.~\ref{fdisp}) yields the following effective parameters: 
\begin{eqnarray} 
(J_{1A},J_{2A},J_{aA},J_{bA}) & = & (144.9{\rm K},95.5{\rm K},11.6{\rm 
K},16.4{\rm K}) \label{esfp} \nonumber \\ (J_{1B},J_{2B},J_{aB},J_{bB}) 
& = & (122.9{\rm K},95.0{\rm K},13.4{\rm K},18.6{\rm K}).
\end{eqnarray}
The dimerization parameters $\lambda_A = 0.659$ and $\lambda_B = 0.772$ 
are rather different in each set of chains, but the values of $J_2$ are 
similar, as expected from the considerations above (V-O-V superexchange). 
These static-model results for the chain couplings $J_1$ and $J_2$ are 
very close to those of the NMR experiment,\cite{rkmyu} and are also in 
accord with values obtained recently by a detailed fit based on 
uncoupled, dimerized chains.\cite{rjsatyu} In comparison with the latter, 
the present procedure has the advantage of containing the true 
dimensionality of the magnon dispersion, so that no discrepancy occurs 
between a gap from a chain fit and a gap including interchain coupling. 
Furthermore, we find that non-crossing magnon dispersion curves have a 
better physical justification due to the similarity of $J_2$ 
superexchange paths in both types of plane. Pressure-dependent 
measurements and comparative {\it ab initio} studies would be required 
to shed further light on the variation of superexchange parameters with 
atomic positions.

In comparison with the single-plane model,\cite{run} the frustrated 
interchain coupling is significantly weaker: in obvious notation, 
$\mu_{aA} = 0.080$, $\mu_{aB} = 0.109$, $\mu_{bA} = 0.113$, $\mu_{bB} 
= 0.151$. Any bound states of the elementary magnons can be expected to 
be at best only very weakly bound, and in addition to have very low 
weight.\cite{run} Another significant difference comes in calculating 
the static susceptibility $\chi(T)$ in the two-plane model: the fit to 
single-crystal susceptibility data\cite{rpbaswll} is much improved in 
terms of $T_{\rm max}$\cite{run} (see Fig.~\ref{fchi}). This arises 
because of the contribution of the higher-energy magnon band. Because 
the calculation of $\chi(T)$ is approximate, in that it involves only 
one-magnon contributions with interaction effects on a mean-field
level, remaining discrepancies between theory and 
experiment are not unexpected. However, this last point cannot 
account for the inconsistency observed in the high-temperature regime, 
where the moments should behave independently, and we return to this 
issue below. 

Two features remain the same as the previous framework. First, we 
contend that two, mutually frustrating AF interactions between the 
alternating chains remain the most likely scenario to account for the 
form of the FM $a$-axis dispersion. This is fitted appropriately by 
the weaker values ($J_a,J_b$) above. Although this may only be 
treated qualitatively, the condition on the Curie temperature, 
\begin{equation}
\Theta_{\rm CW} = - {\textstyle \frac{1}{8}} \sum_{\gamma} [J_{1\gamma} + 
J_{2\gamma} + 2 ( J_{a\gamma} + J_{b\gamma} )], 
\label{ect}
\end{equation}
provides further support for frustrated couplings, because the chain 
couplings alone remain unable to satisfy this sum. Second, the 
two-dimensional dispersion relation in ($q_a,q_b$), with logarithmic 
singularities in the density of states, remains essential to explain 
the temperature scales of thermodynamic properties such as $\chi(T)$ and 
the electron spin resonance absorption.\cite{run}

\section{Spin-Phonon Interactions}

Recent Raman light-scattering experiments on single crystals of 
VOPO\cite{rglgsba} have shown two important new features. One is a strong 
anharmonicity, in the form of a hardening (frequency increase) with 
decreasing temperature, of certain phonons, and the other a strong 
quasielastic scattering of magnetic origin. Both features are clear, 
qualitative evidence for magnetoelastic coupling. In the phonon system, 
the frequencies and polarizations of strongly renormalized phonons 
indicate oscillations of the phosphate (PO$_4$) groups. In particular, 
the 123cm$^{-1}$ phonon observed in $aa$ polarization (in-plane motion 
transverse to the chains, expected\cite{rnkf} to have the strongest 
coupling) loses intensity very rapidly to the spin degrees of freedom. 
In the spin 
system, quasielastic scattering originates in energy fluctuations of 
the spins, which are strongly enhanced by temperature due to phonon 
coupling. In addition, the observed 2-magnon intensity falls rapidly 
with temperature; this feature seems to correspond to the onset of a 
continuum at $\omega = 2\Delta_{0B}$, rather than to a singlet bound 
state as suggested in Ref.~\onlinecite{rglgsba}.

At the qualitative level, this spin-phonon coupling is considerably 
stronger than that observed in the spin-Peierls system CuGeO$_3$. We 
also note here that the authors of Ref. \onlinecite{rglgsba} do not observe 
magnetic scattering at $\omega = 2\Delta_{0A}$, and state that this 
invalidates the 2-plane scenario.\cite{rkmyu} We suggest from $i)$ the 
weakness of the $\omega = 2\Delta_{0B}$ feature and $ii)$ the low 
intensities associated with any calculation of bound states in the 
current parameter regime, that the onset of the second continuum is 
simply too weak to have been detected here. The reported observation 
of this feature in further Raman experiments\cite{rktpbua} verifies 
this hypothesis. 

%\subsection{Theoretical Description}

We wish to provide a theoretical description for the consequences of 
a significant magnetoelastic coupling, in terms of those phonons most 
strongly coupled to a spin system represented by the above, minimal 
model of triplet magnon excitations. Spin-phonon coupling effects 
may readily be envisaged within a conventional, diagrammatic 
treatment,\cite{rnkf} where perturbative inclusion of a magnon-phonon 
vertex with coefficient $g$ would lead to ${\cal O}(g^2)$ self-energy 
corrections to the phonons due to the presence of magnons, and 
conversely. However, this type of approach would appear to be 
precluded here by the difficulties inherent in expressing a 
propagator for hard-core bosons, and in substituting frequency 
summations with constrained thermal occupation functions. This approach 
has been applied to consider spin-phonon coupling in a two-chain 
ladder,\cite{rs} by the introduction of an infinite but fictitious 
repulsion term for triplet excitations on the same bond. In addition 
to this weakness, the bond-operator method appears to be applicable 
at the quantitative level only for very strongly dimerized systems. 
Here we will instead reproduce the mutual renormalization of phonons 
and magnons by a flow-equation method particularly suitable for local 
spin-phonon problems,\cite{rw} in which a unitary transformation 
is applied to the Hamiltonian of the coupled system to eliminate the 
coupling term.   

Motivated by the qualitative observation concerning the importance 
of the PO$_4$ groups, which is further supported by experimental 
observation on the phonon modes involved, in VOPO and related 
compounds,\cite{rglgsba} we will consider the previous model with 
phonon coupling only to the dominant dimer bonds $J_{1A},J_{1B}$. 
We begin with the Hamiltonian in the form 
\begin{eqnarray}
H & = & \sum_{\gamma = A,B} \sum_{i,j} \left\{ J_{1\gamma} 
{\bf S}_{i,j}^1 {\bf .} {\bf S}_{i,j}^2 
 + J_{2\gamma} {\bf S}_{i,j}^2 {\bf .} {\bf S}_{i+1,j}^1 
\right. \label{esph} \nonumber \\ & & \;\;\;\;\;\;\, 
 + \sum_{m = 1,2} \left[ J_{a\gamma} {\bf S}_{i,j}^{m} {\bf .} 
{\bf S}_{i,j+1}^{m}  + J_{b\gamma} {\bf S}_{i,j}^{m} {\bf .} 
{\bf S}_{i,j+1}^{m+1} \right] \nonumber \\ & & \;\;\;\;\;\;\, 
 + \sum_{m = 1,2} J_{f\gamma} {\bf S}_{i,j}^{m} {\bf .} 
{\bf S}_{i+1,j}^{m} \\ & & \;\;\;\;\;\;\, 
\left. + \, \omega_0 b_{i}^{\dag} b_{i} 
 + G (b_i + b_{i}^{\dag}) {\bf S}_{i,j}^1 {\bf .} {\bf S}_{i,j}^2 
\right\}, \nonumber
\end{eqnarray}
where $i$ and $j$ are respectively indices for the dimer bonds ($J_1$) 
along and across the chains in each decoupled plane, and $m = 1,2$ 
denotes the left or right spin in each dimer. The phonons $\{b_i\}$ 
are introduced as local, Einstein modes 
of fixed frequency $\omega_0$ at each dimer bond $i$; in reciprocal space 
these are nondispersive, and have the same weight at all wave vectors 
${\bf q}$. Because these phonon modes involve motion of the PO$_4$ groups, 
and not of the magnetic (V$^{4+}$) ions themselves, correlations between 
displacements may safely be neglected, and the approximation of Einstein 
phonons justified. The spin-phonon coupling constant $G$ is a free 
parameter to be fixed from experiment, but may in principle be very 
large: comparison with CuGeO$_3$ suggests that values exceeding 
0.3$J_{1A}$ are not excluded.\cite{rwgb} The term in $J_{f\gamma}$ 
describes a frustrating, next-neighbor coupling along the dimerized 
spin chains; this is zero in the bare model (Sec. II), but is generated 
at second order in $G$ by the unitary transformation which eliminates 
the final term to leave only phonon terms bilinear in $\{b_i\}$ in the 
resulting, effective Hamiltonian. A systematic discussion of the 
transformation procedure is presented in Ref.~\onlinecite{ru2}.

In deriving the effective Hamiltonian we retain only the leading order 
in $G/\omega_0$, which is $(G/\omega_0)^2$; omission of next-order 
terms $(G/\omega_0)^4$ can be expected to be well justified. The 
transformation also involves an expansion in $J/\omega_0$, where terms 
${\cal O}(J/\omega_0)$ are retained but those 
${\cal O}((J/\omega_0)^2)$ omitted; the 
validity of this approximation is not apparent in VOPO, which from the 
values of $J_{1\gamma}$ and $\omega_0$ is not well in the anti-adiabatic 
limit, but is motivated by the good, semi-quantitative agreement with 
simulations for a similar model by Bursill {\it et al.},\cite{rbmh} 
and may also be justified {\it a posteriori}. 

The general form of the effective Hamiltonian may be represented as 
\begin{equation}
H = H_{\rm spin} + H_{\rm phonon} + \Delta H_X + \Delta H_Y + \Delta H_Z ,
\label{egfeh}
\end{equation}
in which $H_{\rm spin}$ and $H_{\rm phonon}$ denote respectively the pure 
spin and phonon parts of Eq. (\ref{esph}). The coupling term is transformed 
into two correction terms in the spin sector, which following the notation 
of Ref.~\onlinecite{ru2} 
we write as 
\begin{equation}
\Delta H_X = - \frac{1}{\omega_0} \sum_{i,j} A_{i,j}^{\dag} A_{i,j} 
\label{edhx}
\end{equation}
and 
\begin{equation}
\Delta H_Y = \frac{1}{2 \omega_0^2} \coth \left( \frac{\omega_0}{2T} \right) 
\sum_{i,j} \left[ A^{\dag}_{i,j}, {\cal L} A_{i,j} \right].
\label{edhy}
\end{equation}
Here $A_{i,j}$ denotes the local coupling, which we have taken as $G
{\bf S}_{i,j}^1 {\bf .} {\bf S}_{i,j}^2$,\cite{footnote} and ${\cal LA} 
= [H_S,{\cal A}]$ denotes the commutator of the quantity ${\cal A}$ with 
the spin-only part $H_S$ of the starting Hamiltonian (Eq. (\ref{esph})).
A further correction term arises for the phonon sector, 
\begin{equation}
\Delta H_Z = \frac{1}{\omega_0^2} \sum_{i,j} b_{i,j}^{\dag} b_{i,j} 
\langle \left[ A^{\dag}_{i,j}, {\cal L} A_{i,j} \right] \rangle_S,
\label{edhz}
\end{equation}
where the expectation value $\langle \dots \rangle_S$ is computed for the 
spin sector. This last contribution was not considered in 
Ref.~\onlinecite{ru2}, where the focus was on the spin sector, and arises 
in the phonon sector from a mean-field treatment of the terms in Eq.~(9e) 
of that work. 

Explicit evaluation of the additional terms generated by the transformation 
(Eqs. (\ref{edhx}-\ref{edhz})) yields 
\begin{equation}
\Delta H_X = - \frac{G^2}{\omega_0} \sum_{i,j} \left( {\bf S}_{i,j}^{1} 
{\bf .S}_{i,j}^2 \right)^2 \, = \, \frac{G^2}{2 \omega_0} \sum_{i,j} 
\left( {\bf S}_{i,j}^{1} {\bf .S}_{i,j}^2 \right),
\label{edhxw}
\end{equation}
to within a constant, using that for spins $S = 1/2$, $\left({\bf S}_1 
{\bf .S}_2 + 3/4 \right)^2 = {\bf S}_1 {\bf .S}_2 + 3/4 $ because the 
eigenvalues of the right-hand side are 0 and 1. $\Delta H_Y$ and 
$\Delta H_Z$ are calculated from the result
\begin{eqnarray}
\left[ A^{\dag}_{i,j}, {\cal L} A_{i,j} \right] \! & = & {\textstyle 
\frac{1}{2}} G^2 \left\{ J_2 \left( {\bf S}_{i+1,j}^1 - {\bf 
S}_{i-1,j}^2 \right) \left( {\bf S}_{i,j}^1 - {\bf S}_{i,j}^2 \right) 
\right. \label{ecala} \nonumber \\ & & - (J_a - J_b) \left( 
{\bf S}_{i,j}^1 - {\bf S}_{i,j}^2 \right) \\ & &  
\left. \times \! \left[ \left( {\bf S}_{i,j+1}^1 \! - \! {\bf S}_{i,j+1}^2 
\right) \! + \! \left( {\bf S}_{i,j-1}^1 \! - \! {\bf S}_{i,j-1}^2 
\right) \right] \right\}, \nonumber 
\end{eqnarray}
where four-spin terms are neglected. With due attention to the number 
of neighboring chains, the additional terms $\Delta H_X$, $\Delta H_Y$, 
and $\Delta H_Z$ may be transcribed into the form of the initial 
Hamiltonian (\ref{esph}), as corrections to the couplings $\{J\}$, 
\begin{mathletters}
\label{eccc}
\begin{eqnarray}
\Delta J_{1\gamma} & = & \frac{G^2}{2 \omega_0} \\
\Delta J_{2\gamma} & = & - \frac{G^2}{2 \omega_0^2} J_{2\gamma} \coth 
\left( \frac{\omega_0}{2T} \right) \\
\Delta J_{f\gamma} & = & \frac{G^2}{4 \omega_0^2} J_{2\gamma} \coth 
\left( \frac{\omega_0}{2T} \right) \\
\Delta J_{a\gamma} & = & - \frac{G^2}{2 \omega_0^2} \left( 
J_{a\gamma} - J_{b\gamma} \right) \coth \left( \frac{\omega_0}{2T} \right) 
\\ \Delta J_{b\gamma} & = & \frac{G^2}{2 \omega_0^2} \left( 
J_{a\gamma} - J_{b\gamma} \right) \coth \left( \frac{\omega_0}{2T} \right), 
\end{eqnarray}
\end{mathletters}
and as a correction to the phonon frequency in each plane of 
\begin{equation}
\Delta \omega_{0\gamma} = \frac{G^2}{\omega_0^2} J_{2\gamma} \left( 
\langle {\bf S}_i^1 {\bf .S}_{i+1}^1 \rangle - \langle {\bf S}_i^2 
{\bf .S}_{i+1}^1 \rangle \right).  
\label{epfc}
\end{equation}
The latter result was obtained by neglecting additional interchain 
coupling terms, on the grounds that both the coupling constants 
$|J_a - J_b| \ll J_2$ and the spin expectation values are much smaller 
between the chains than within them. The expectation values in Eq. 
(\ref{epfc}) may be computed for independent chains by a high-temperature 
expansion in combination with the known $T = 0$ behavior,\cite{rbeu,rblu} 
as described in App.~A.

These values are all finite at zero temperature. In order to perform a 
self-consistent calculation, we deduce the appropriate bare values of 
the coupling constants such that their renormalized values correspond 
to the experimental data. Thus the fitting parameters provided in the 
previous section are effective quantities already containing the low-$T$ 
renormalization. We now discuss the quantitative properties of these 
corrections in the context of phonon anharmonicity, magnon thermal 
renormalization, and thermodynamic properties.

\section{Experimental consequences}

\subsection{Phonon anharmonicity}

One observes from Eq. (\ref{epfc}) that $\Delta \omega_0 > 0$, because 
for an AF system $\langle {\bf S}_i^2 {\bf \cdot S}_{i+1}^1 \rangle$ is 
negative. Further, one expects that $\Delta \omega_0 
\rightarrow 0$ as temperature becomes very large and the expectation 
values vanish. Both of these features are consistent with experiment. 
More interestingly still, $\Delta \omega_0$ does not vary monotonically 
with temperature, as $\langle {\bf S}_i {\bf .S}_j \rangle$ for nearest 
neighbors increases in magnitude at low temperatures; this result is 
exactly in accord with the small, initial increase observed\cite{rglgsba} 
with temperature for both 70cm$^{-1}$ and 123cm$^{-1}$ phonons. 

\begin{figure}[hp]
\centerline{\psfig{figure=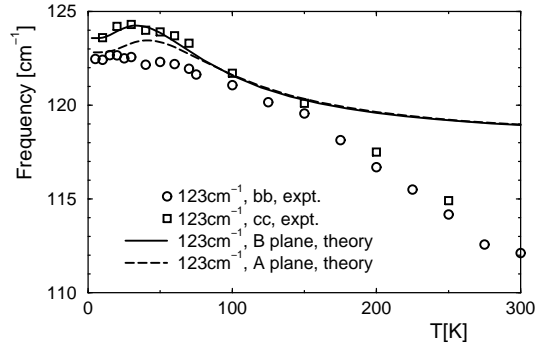,height=5cm,angle=270}}
\medskip
\caption{\label{fom}
Thermal renormalization due to magnons, for a phonon of  
frequency $\omega_0 = 118.4$cm$^{-1}$ with coupling $G = 6.6$meV. 
Symbols are experimental data in $cc$ and $bb$ polarizations, from Ref. 
\protect\onlinecite{rglgsba}.} 
\end{figure}

Fig.~\ref{fom} shows the thermal renormalization effect for an Einstein 
phonon of  bare frequency $\omega_0 = 118.4$cm$^{-1}$ (14.68meV). 
The bare frequency value is chosen such that the effective frequency
including corrections coincides with the experimental value of 123cm$^{-1}$ 
at $T=0$. The solid and dashed 
lines correspond to the different effects of the magnon modes in each 
type of plane for the same spin-phonon coupling. The coupling constant 
$G = 6.6$meV was chosen to reproduce the initial upturn seen in $cc$ 
polarization. Note that we do not wish to imply a connection between 
the two plane types and the two polarizations: each polarization 
measures some combination of the contributions from each plane type, 
determined by possibly different 
coupling constants to each. However, the results in Fig.~\ref{fom} provide a 
good indication of general magnetoelastic coupling effects, and of the 
splitting in energy of a single phonon mode when further couplings are 
neglected, which from comparison with experiment appears realistic. For 
completeness, we add here that the results in Fig.~\ref{fom} were 
computed with the bare coupling constants
\begin{eqnarray} 
(J_{1A},J_{2A},J_{aA},J_{bA}) & = & (132{\rm K},112{\rm K},10.9{\rm 
K},17.1{\rm K}) \label{ebsfp} \nonumber \\ (J_{1B},J_{2B},J_{aB},J_{bB}) 
& = & (111{\rm K},110{\rm K},12.5{\rm K},19.5{\rm K}).
\end{eqnarray}
Comparison with (\ref{esfp}) shows that the coupling $G$ has a  
strong effect on the chain superexchange parameters $J_1$ and $J_2$, 
but does not alter strongly the interchain values $J_a$ and $J_b$, 
whose sum we have determined as in Sec. II. 

The most obvious qualitative feature in Fig.~\ref{fom} is that magnon 
effects alone cannot account for the observed anharmonicity over the 
full temperature range. This appears to consist of two components: 
at low temperatures, magnon-related renormalization effects occur on 
the energy scale of the magnon gap $\Delta_{0B}$ = 35K; at higher 
temperatures, a further decrease of $\omega_0$ is present, on the 
energy scale (180K) of $\omega_0$ itself. The latter behavior we 
ascribe to the lattice structure, and await dynamical simulations, 
measurements of the thermal expansion coefficients, and magnetostriction 
experiments to verify this point. Recent measurements\cite{rwsszl} of 
ultrasound propagation in (VO)$_2$P$_2$O$_7$ performed in high, pulsed 
fields found a strong magnetoacoustic coupling, and indicate that the 
required lattice information may be readily extracted. The small maximum 
in the phonon frequency at low $T$ would not be expected from lattice 
anharmonicity, and the explanation in terms of a magnon renormalization 
justifies its use in fitting the coupling constant. The resulting value 
of 0.58 for $g = G/J_{1A}$ ($J_{1A}$ taken from Eq.~(\ref{ebsfp})) 
reflects the very strong coupling present in 
VOPO. It is not implausible in view of the complicated superexchange path 
and its demonstrated sensitivity to the atomic position of P, and not 
inconsistent with expectations based on a comparison with CuGeO$_3$. We 
note finally that all ${\bf q}$-dependence of the phonon modes has been 
omitted from our discussion, for the reasons specified in the previous 
section. 

\subsection{Magnon renormalization}

Inspection of the renormalized coupling constants (Eqs. (\ref{eccc})) 
reveals the following qualitative features. $i)$ the strong interaction 
constant $J_1$ has a temperature-independent enhancement; $ii)$ the weaker 
bond $J_2$ is suppressed with increasing temperature; $iii)$ in-chain 
frustration $J_f$ is introduced, and rises with increasing $T$; $iv)$ the 
interchain coupling diminishes as temperature rises. Based on these 
observations, one expects a band flattening in the chain direction: 
the average frequency depends mostly on $J_1$, and is little changed, 
whereas the difference between the extrema, $\omega_{\rm max} - 
\omega_{\rm min}$ depends mostly on $J_2$ (and partly on $J_f$), and 
decreases with increasing temperature.

\begin{figure}[hp]
\centerline{\psfig{figure=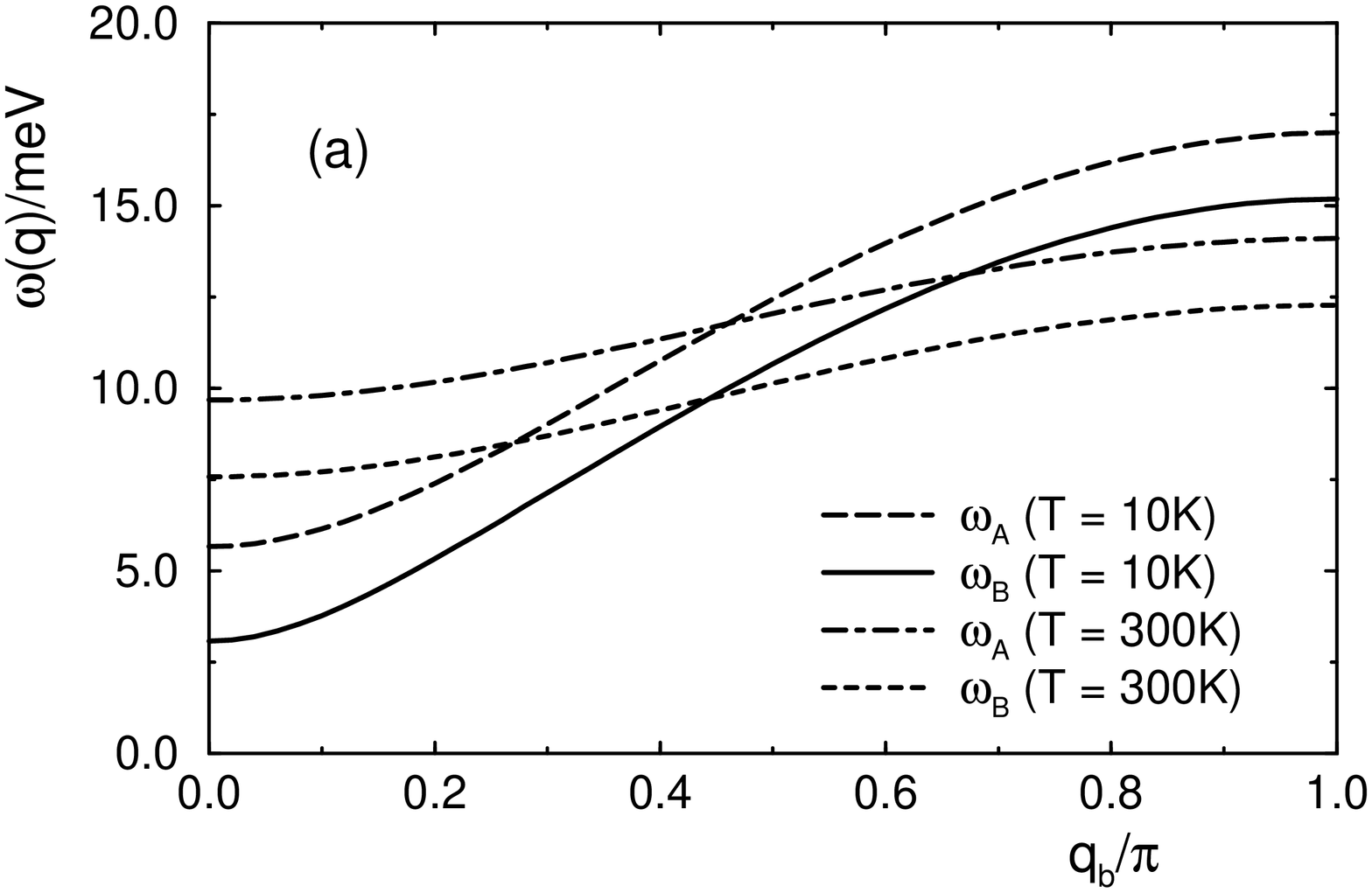,height=5cm,angle=270}}
\centerline{\psfig{figure=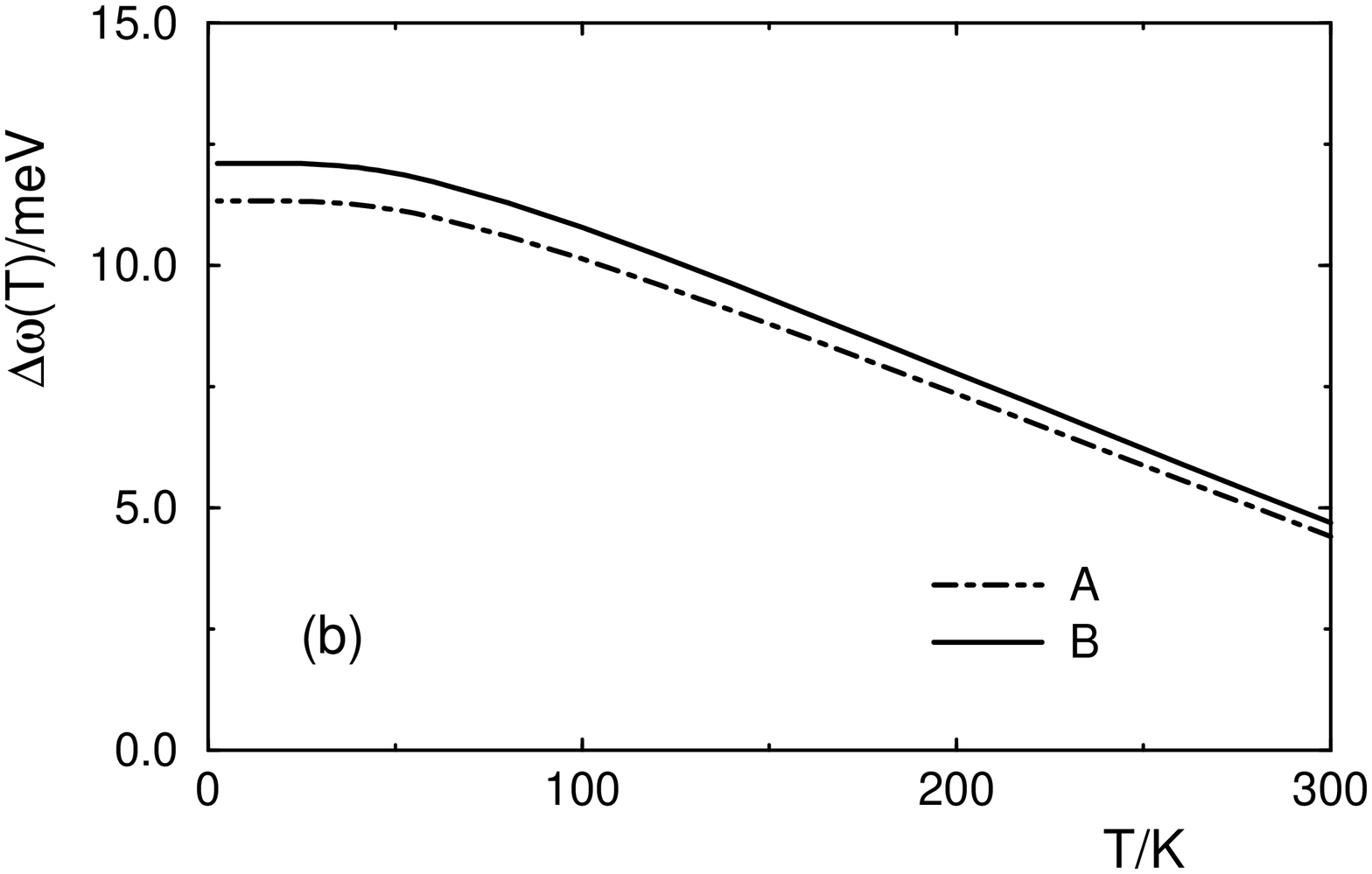,height=5cm,angle=270}}
\medskip
\caption{\label{fE1}
(a) Thermal (phonon) renormalization of elementary magnon modes at 
$q_a = 0$, illustrated for an Einstein phonon with frequency $\omega_0 = 
118.4$cm$^{-1}$ and coupling $G = 6.6$meV, at temperatures of 10K 
and 300K. (b) Band-flattening effect of thermal renormalization, shown 
by thermal evolution of $\Delta \omega = \omega(0,\pi) - \omega(0,0)$,
the difference between dispersion extrema. }
\end{figure}

Fig.~\ref{fE1}(a) illustrates this effect by comparing the magnon dispersion 
relations in the chain direction, determined at low and high 
temperatures for a spin system coupled to an Einstein phonon mode 
at $\omega_0 = 118.4$cm$^{-1}$, with the same spin-phonon coupling, 
$G = 6.6$meV, determined from the anharmonicity. Flattening of the 
magnon bands as a consequence of this phonon coupling is clearly 
very strong. This result leads to a readily falsifiable prediction: 
INS measurements could be expected to reveal large upward shifts in 
the magnon dispersion peak position at the zone centre, and downward 
shifts at the zone boundaries, even at temperatures where broadening 
of the signal does not compromise the fit quality. Our prediction
for this effect is shown as a function of temperature in Fig.~\ref{fE1}(b). 
Finally, we note that similar, very strong band-flattening 
features have been observed in INS measurements performed over a 
range of temperatures for the material KCuCl$_3$, where complicated 
superexchange paths are also known to be important in determining 
the magnetic couplings.\cite{rcu}

\subsection{Thermodynamic quantities}

\begin{figure}[hp]
\centerline{\psfig{figure=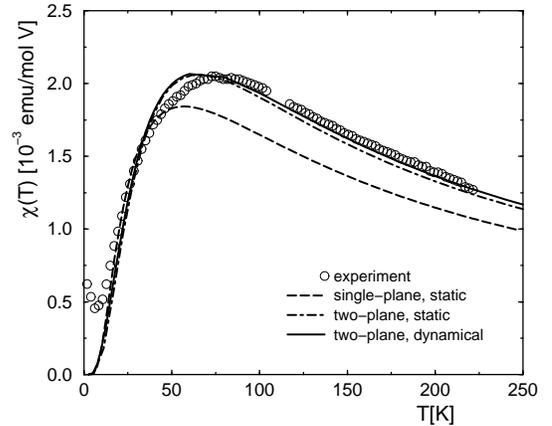,height=6.5cm,angle=270}}
\medskip
\caption{\label{fchi}
Comparison of experimental susceptibility $\chi(T)$ (circles) 
with single-plane model (dashed), static, two-plane model (dot-dashed), 
and dynamical, two-plane model containing renormalizing effects of 
an Einstein phonon with frequency $\omega_0 = 118.4$cm$^{-1}$ and 
coupling $G = 6.6$meV (solid). Experimental data from 
Ref.~\protect\onlinecite{rpbaswll}. } 
\end{figure}

The magnon thermal renormalization arising as a result of spin-phonon 
coupling should be manifest in corrections to thermodynamic quantities 
at high temperatures. Considering first the static susceptibility 
$\chi(T)$,\cite{rpbaswll} this may be computed\cite{run} by integration 
over the available magnon modes. $\chi(T)$ grows initially with temperature, 
its behavior governed exponentially by the gaps, has a maximum on the 
order of the dominant energy scales $J_1$, and converges at high-$T$ to 
Curie-Weiss behavior. Fig.~\ref{fchi} compares the single-crystal experimental 
susceptibility with the single-plane model predictions\cite{run}, the 
two-plane, static model for the parameters fitted in Eqs. (\ref{esfp}), 
and the two-plane, dynamical model with thermally renormalized magnon 
dispersions (Eqs. (\ref{ebsfp})). In the last case, one Einstein phonon 
mode per bond ($\omega_0 = 118.4$cm$^{-1}$, $G = 6.6$meV) is used, as in 
the calculations of the previous section. 

As noted in Sec. II, the two-plane, static model gives a rather good 
account of the low-temperature regime and the maximum, and remains 
quite satisfactory at higher temperatures. It is perhaps surprising 
in view of the large coupling constant $G$ that the effects of dynamical 
phonons are so small. However, it is also evident that the model including 
thermal renormalization of the superexchange parameters provides excellent 
agreement in all three temperature regimes, including high-$T$. 
This result not only confirms the consistency of the method, and of the 
value of $G$ deduced from the phonon anharmonicity, but has a profound 
consequence for interpretation of high-$T$ susceptibility data in terms 
of a Curie-Weiss temperature in the presence of phonons. The effects of a 
spin-phonon coupling on $\Theta_{\rm CW}$ may be computed rigorously, 
as shown in App.~B. It is found that the value of $\Theta_{\rm CW}$ 
including spin-phonon coupling corresponds to that given in Eq.~(\ref{ect}) 
if the renormalization of $J_{1\gamma}$ due to $\Delta H_X$ (\ref{edhx}) 
is taken into account. Corrections due to $\Delta H_Y$ (\ref{edhy}) are 
found to cancel. We note finally that in the real material, several types 
of phonon mode contribute to a susceptibility renormalization. However, 
because their effects are proportional to the square of the corresponding 
coupling constant, contributions beyond that from the dominant phonon 
mode may be weak.

\begin{figure}[hp]
\centerline{\psfig{figure=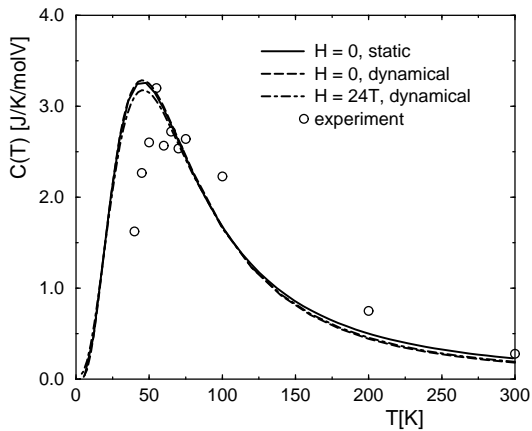,height=6.5cm,angle=270}}
\medskip
\caption{\label{fC}
Specific heat of magnetic system, comparing static (solid) and 
dynamical (dashed) two-plane models. Dot-dashed line is for dynamical 
model in a magnetic field of 24T. Points are experimental data from 
Ref.~\protect\onlinecite{rglgsba}. Dynamical models contain one Einstein 
phonon of frequency $\omega_0 = 118.4$cm$^{-1}$ and coupling $G = 6.6$meV. } 
\end{figure}

We close this section by computing a further experimentally accessible 
quantity which is expected to illustrate thermal renormalization due 
to magnetoelastic coupling. The specific heat of the spin system alone 
is difficult to isolate from phonon contributions, and to date has been 
deduced only indirectly from quasielastic Raman scattering.\cite{rglgsba}
The calculation proceeds as for the magnetic susceptibility,\cite{run} 
but with integration in reciprocal space over the magnon contributions to 
the second temperature derivative of the free energy, and may be readily 
extended to finite magnetic fields. The results presented in Fig.~\ref{fC} 
demonstrate again both the utility of the static model and the small 
but significant upward renormalization offered by the dynamical model 
at high $T$. Both sets of data are close to the results of 
calculations\cite{rrk} for a single dimerized chain with appropriate 
$\lambda$, indicating that interchain coupling effects have only a small 
effect on the overall features of thermodynamic quantities. By contrast, 
a strong magnetic field is most effective at low $T$, where it acts to 
reduce the gap, and at intermediate $T$ where the peak value of $C(T)$ 
is suppressed. The experimental data show too much scatter to be of 
particular utility, and we await more refined measurements of this quantity. 

\section{Summary}

In conclusion, we have analyzed a two-plane model for VOPO, which is 
fully consistent with the known crystal structure, and emphasize the 
key role of the PO$_4$ groups mediating the dominant superexchange path. 
A full understanding of the magnetic properties is not possible 
without considering the strong magnetoelastic coupling effects observed 
in Raman light-scattering experiments. We account for these 
qualitatively in terms of phonon modes of the phosphate groups, 
and quantitatively by using the flow-equation method. Phonon hardening 
is found to have contributions from both spin coupling and lattice 
anharmonicity, although only the former can account for the observed 
low-temperature softening, which provides an estimate of the spin-phonon 
coupling constants. 

We obtain an excellent fit to the magnon dispersion data for the two 
triplet modes observed by INS, and find complete agreement with the 
parameters deduced from NMR studies. In addition, there is a strong 
thermal renormalization of the magnon dispersions due to phonons, in 
the form of a band flattening, which should be clearly visible in INS 
measurements performed as a function of temperature. We show further  
that, while the static two-plane model parameters give a good account 
of static susceptibility and specific heat data, the renormalization 
effects of dynamical phonons lead to additional corrections which are 
required to reproduce the high-temperature limit. However, we have found 
that even for very strong coupling, the influence of phonons on 
thermodynamic magnetic properties is rather small, and such quantities 
may be quite well described by a static model. Unambiguous evidence of 
phonon effects is provided only by dynamical properties, such as Raman 
spectra and triplet dispersions.

\section{Acknowledgements}

We are grateful to A. B\"uhler, M. Enderle, P. Lemmens, U. L\"ow, 
S. Nagler and H. Schwenk for helpful discussions and provision of data. 
This work was supported by the Deutsche Forschungsgemeinschaft, through 
SP 1073 and SFB 341 (GSU), and through SFB 484 (BN).

\appendix

\section{Dlog-Pad\'e extrapolation of expectation values}

\begin{figure}[hp]
\centerline{\psfig{figure=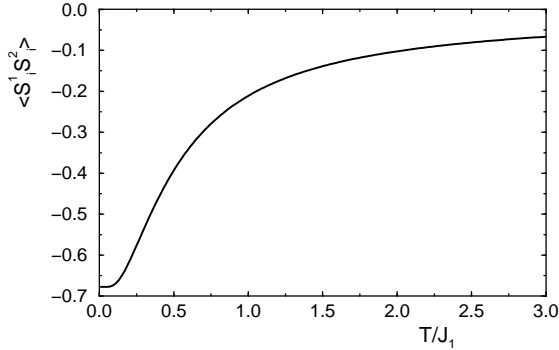,width=8cm,angle=270}}
\medskip
\caption{Expectation value of strong-bond spin correlation function 
$\langle {\bf S}_i^1 {\bf \cdot} {\bf S}_i^2\rangle$, determined as  
function of temperature for $J_2/J_1 = 0.75$ and $J_f/J_1 = 0.05$ by 
procedure described in text.\label{extrapol}}
\end{figure}

In this appendix we present details of the computation of expectation 
values such as $\langle {\bf S}_{i,j}^1 {\bf \cdot} {\bf S}_{i,j}^2 
\rangle$, as functions of temperature. The basic physical consideration 
is  straightforward for a gapped spin system, and we will demonstrate 
it for the expectation value $A(T)= \langle {\bf S}_i^1 {\bf \cdot} 
{\bf S}_i^2\rangle$ of the spin correlation function for the 
strong bond in a dimerized spin chain ($i$ is the dimer index).
At zero temperature, $A(T)$ has the finite, negative value $A_0$. 
At small but finite temperature, $A(T)$ will deviate from $A_0$, with an 
exponentially small deviation due to the presence of the spin gap. 
It is plausible to take this deviation to be positive, $A(T) - A_0 \ge 0$, 
and so one expects ({\it cf}.~Fig.~\ref{extrapol})
\begin{mathletters}
\begin{eqnarray}\label{asym}
A(T) &\approx & A_0 + A_1 T^\nu \exp(-\Delta/T)\ \mbox{for}\ T\ 
\mbox{small}\\ A(T) &\propto & 1/T \ \mbox{for}\ T\ \mbox{large},
\end{eqnarray}
\end{mathletters}
where $\Delta$ is the spin gap. The exponent $\nu = d/2$ is determined
for a dispersion with quadratic minima only by the dimensionality $d$.
For the quantitative computation we use two sources of input, namely 
$i)$ an expansion at zero temperature about the dimer limit, such as that 
given in Ref.~\onlinecite{knett00a}, and $ii)$ a high temperature 
expansion such as that presented in Refs.~\onlinecite{rbeu,rblu}.

From $(i)$ one may deduce  the values of $A_0$ and $\Delta$. For a 
sufficiently dimerized system such as (VO)$_2$P$_2$O$_7$, the plain 
truncated series can be used. The value of $A_0$ is obtained from the
expansion of the ground state energy as a function of the strong-bond 
coupling $J_1$, the weak-bond coupling $J_2$, and the next-nearest-neighbor
coupling $J_f$, by partial differentiation with respect to $J_1$.

The extrapolation is performed using a Dlog-Pad\'e approximant for 
$A(T) - A_0$ in the following manner. Expansion in $\beta$ up to 
9$^{\rm th}$ order is used to describe 
\begin{equation}
a:=\partial_\beta A(1/\beta)/(A(1/\beta)-A_0) .
\end{equation}
Then, because we expect regular behavior in the limit $\beta \to \infty$, 
it is appropriate to change to variable $u = \beta/(1+\beta)$ (whence  
$\beta = u/(1-u)$). Because $u\propto \beta$ for small values of $u$ and 
$\beta$, an expansion of $a$ in terms of $u$ is available up to order
8. This is used to set up a $[8,2]$ Pad\'e approximant $P_8/Q_2$ in $u$,
where the subscripts stand for the degree of the polynomial in $u$. The 
missing constants for order 9 and 10 are determined from the known 
asymptotic behavior. From Eq. (\ref{asym}) we have $a = -\Delta - 
d/(2\beta)+{\cal O}(\beta^{-2})$, which yields 
\begin{eqnarray}
P_8/Q_2\Big|_{u=1} &=& -\Delta\\
\partial_u (P_8/Q_2)\Big|_{u=1}&=& 1/2
\end{eqnarray}
for chain geometry. In this way, reliable and non-defective interpolations
between the zero-temperature result and the high-temperature behavior can 
be obtained, as illustrated in Fig.~\ref{extrapol}.

Applying the above procedure to the phonon frequency shift (\ref{epfc}),
one encounters a robust pole in the Pad\'e approximants. This pole
is approximately independent of the actual Pad\'e approximant, so 
is not spurious. In fact the procedure explained above must fail if 
$A(T) - A_0$ changes sign, an occurrence which may not be excluded for
general expectation values. Indeed, we find that the weak-bond expectation
value in a dimerized chain does display such a sign change, although 
the difficulties this introduces are readily avoided. The function 
considered is separated into two parts, each without sign change. In practice, 
this was effected by adding to the expectation values in Eq.~(\ref{epfc}) 
four times the strong-bond expectation value, and applying the above 
procedure to the combined expression. The strong-bond contribution from 
separate computation (above) is then subtracted to yield the final result.

This technique, which relies on a physically appropriate combination of 
information from zero and high temperatures, permits rapid computation of 
the frequency shift despite the need to determine the effective couplings 
anew at each temperature.

\section{Curie-Weiss temperature and spin-phonon coupling}

In this appendix we derive the effect of a magnetoelastic coupling on the 
Curie-Weiss temperature, $\Theta_{\rm CW}$. Because at high temperatures 
one assumes $\chi(T) \propto (T- \Theta_{\rm CW})^{-1}$, this procedure 
involves computing the subleading $1/T$ behavior in $(T- \Theta_{\rm 
CW})^{-1} = 1/T + \Theta_{\rm CW}/T^2 + {\cal O}(T^{-3})$. We consider 
first a uniform spin chain coupled to Einstein phonons,
\begin{equation}
H = J\sum_i {\bf S}_{i} {\bf \cdot}{\bf S}_{i+1}(1+g(b_i+b_i^\dagger))
+\omega b^\dagger_i b_i \ .
\end{equation}
Direct expansion in $1/T$ of the partition function ${\rm Tr}[ \exp(-\beta 
H)]$ is not possible due to the infinite-dimensional bosonic Hilbert space,
and we employ instead an expansion in $J$.

The first two terms read
\begin{eqnarray}\nonumber
{\rm Tr} [\exp(-\beta H)]{} &=&{\cal O}(J^3) + Z_0^N{\bf \cdot}\\ 
\label{expan} &&\hspace{-2.5cm}\left(1 + N J^2 
{\rm Tr_{spins 1,2}}[({\bf S}_1{\bf \cdot}
{\bf S}_2)^2](\beta^2/2! + g^2 f_1/Z_0) \right) ,
\end{eqnarray}
with $Z_0 = (1-e^{-\beta\omega})^{-1}$ the bosonic partition function 
and the coefficient $f_1$ given by 
\begin{eqnarray}\nonumber
f_1 &=& \int_0^\beta  \int_0^\beta \int_0^\beta d\beta_1 d\beta_2 d\beta_3
 \delta(\beta-\beta_1-\beta_2-\beta_3) \cdot
\\ \label{fcoeff}
&&\hspace{1cm}
{\rm Tr_{boson}} \left[e^{-\beta_1\omega b^\dagger b}u e^{-\beta_2\omega 
b^\dagger b} u e^{-\beta_3\omega b^\dagger b}\right] \ ,
\end{eqnarray}
where $u$ denotes $b + b^\dag$. The coefficients $f_m$ are
the generalization of (\ref{fcoeff}) to $2m$ $\beta$-ordered factors of $u$.
The evaluation of ${\rm Tr_{spins 1,2}}({\bf S}_1{\bf \cdot}{\bf S}_2)^2$ 
is straightforward, and yields $3/16$. The evaluation of $f$ is possible 
by explicit calculation, or more easily by the observation that $f$ equals
the second coefficient in $\lambda$ in an expansion of $Z(\lambda)$, where 
\begin{equation}
Z(\lambda) = Z_0 + \sum_m \lambda^{2m} f_m \ .
\end{equation}
is the partition function of $H = \omega b^\dag b + \lambda u$. 
Transforming to shifted bosons $\tilde b = b + \lambda/\omega$, the shifted
partition sum is easily found to be $Z(\lambda) = Z_0\exp(\beta\lambda^2 / 
\omega)$, whence it follows directly that $f_m = Z_0(\beta/\omega)^m/m!$.
Because $f_m$ is of order $\beta^m Z_0$, the general phononic contribution 
in Eq.~(\ref{expan}) is of order $J^{2m}(\beta/\omega)^m$, or higher in 
$\beta$. Hence one obtains a systematic high-temperature expansion 
circumventing the problem of the infinite dimensional bosonic Hilbert 
space, which is cut off by the factor $e^{-\beta\omega b^\dagger b}$.

In analogy to the expansion (\ref{expan}), the variance of the 
magnetization $M:=\sum_i S^z_i$ is given by 
\begin{equation}
  \frac{1}{N} {\rm Tr} \left[M^2\exp(-\beta H)\right] = \frac{1}{4}R_1+R_2 +
{\cal O}(J^3) , 
\label{expan2}
\end{equation}
where $R_1$ is the partition sum (\ref{expan}) and 
\begin{eqnarray}\nonumber
  R_2 &=& 2Z_0^N\left(-\beta J{\rm Tr_{spins 1,2}}\left[S^z_1S^z_2
      ({\bf S}_1{\bf \cdot}{\bf S}_2)\right]  +\right.\\
  &&\hspace{-1cm}\left. J^2 {\rm Tr_{spins 1,2}}\left[S^z_1S^z_2
      ({\bf S}_1{\bf \cdot}
      {\bf S}_2)^2\right]\left(\beta^2/2+g^2 f_1/Z_0\right)\right)\ .
\end{eqnarray}
The susceptibility $\chi(T)$ is the ratio of Eqs.~(\ref{expan2}) and 
(\ref{expan}) to order $J^2$, 
\begin{eqnarray}
4T\chi & = & 1 - \frac{\beta J}{2} - \frac{\beta^2 J^2}{8} - \frac{\beta 
g^2J^2}{4\omega} +{\cal O}(J^3\beta^2)\\ & = & 1 - \beta\left(\frac{J}{2} 
+ \frac{g^2 J^2}{4\omega} \right) + {\cal O}(\beta^2) ,
\end{eqnarray}
from which follows
\begin{equation}
\label{result}
\Theta_{\rm CW} = - \left(\frac{J}{2} + \frac{g^2J^2}{4\omega} \right) .
\end{equation}
This result implies quite generally that the Curie-Weiss temperature
is lowered by spin-phonon interactions, {\it i.e.} in AF systems the 
modulus of $\Theta_{\rm CW}$ increases. For Einstein phonons the 
contribution of spin-phonon coupling to the Curie-Weiss temperature 
amounts to $-G^2/(8\omega)$ per bond linked to each site with coupling 
$G = gJ$. This is exactly the same effect as the renormalization 
contained in $\Delta H_X$ (Eq.~(\ref{edhx})), a fact which corroborates
the validity and utility of the flow-equation approach.

To conclude this appendix, we have for the model (\ref{esph}) a change 
of $\Theta_{\rm CW}$ due to spin-phonon interactions given by 
\begin{equation}
\Theta_{\rm CW; total} = \Theta_{\rm CW; bare} - G^2/(8\omega)
\end{equation}
where $\Theta_{\rm CW; bare}$ is the value given in Eq.~(\ref{ect}).
The expansions of Eqs.~(\ref{expan},\ref{expan2}) provide an interesting 
and systematic extension to higher orders in $J$. Generally, a calculation 
in $J^{2m}$ will provide results for a high-temperature expansion up to 
$\beta^m$. 


\begin{thebibliography}{99}
\bibitem{fn} Note that the coefficients $t_{ij}$ presented in 
Ref.~\onlinecite{run} contain two minor errors: in $t_{20}$, 
$- \lambda^2 / 16$ should read $- \lambda^2 / 32$, and in $t_{11}$, 
$\lambda \mu^2_- / 8$ should read $\lambda \mu_- / 8$.
\end{thebibliography}
\end{document}